\def   \ni {\noindent}

\def   \ssk {\vskip  5truept}

\def   \bsk {\vskip 15truept}
 
\def   \newpage {\vfill\eject}
\def   \newline {\hfil\break}

\documentstyle[]{article}
\input{epsf}
\begin{document}

\hsize 5truein
\vsize 8truein
\font\abstract=cmr8
\font\keywords=cmr8
\font\caption=cmr8
\font\references=cmr8
\font\text=cmr10
\font\affiliation=cmssi10
\font\author=cmss10
\font\mc=cmss8
\font\title=cmssbx10 scaled\magstep2
\font\alcit=cmti7 scaled\magstephalf
\font\alcin=cmr6 
\font\ita=cmti8
\font\mma=cmr8
\def\ref{\par\noindent\hangindent 15pt}
\null


\title{\ni 
Restrictions on the parameters of boundary layer and accretion disk of X-ray
bursters in the low state}                                               

\bsk \bsk
\author{\ni 
Revnivtsev M.$^{1,2}$, Gilfanov M.$^{2,1}$, Churazov E.$^{2,1}$, Sunyaev R$^{2,1}$.
}                                                       
\bsk
\affiliation{
Space Research Institute, Moscow, Russia\\
Max-Planck Institut f\"ur Astrophysik, Garching, Germany
}                                                
\bsk
\baselineskip = 12pt

\abstract{ABSTRACT \ni
Using ASCA observations of several X--ray bursters in the low
spectral state ($L_x\sim$ 1-5 $\times10^{36}$ erg/s) restrictions 
on the parameters of the accretion disk  
and the boundary layer were obtained. The low state spectra of X--ray
bursters observed with ASCA often can be well described by a power law
with photon index of 1.7--2.2  and are not consistent with
significant contribution of the soft spectral component to the total
luminosity. For a blackbody spectrum with temperature of 0.2--2 keV
the upper limits correspond to $<$ 10--20\% of the total 0.5--10
keV luminosity. On the other hand, theoretical calculations
predict that in the case of a Keplerian accretion disk around a slowly
rotating neutron star (with radius of 3$R_g$) $\sim$2/3 of the  total
gravitational energy released in the system can be released in the
boundary layer between the inner part of the accretion disk and the
neutron star surface. 
 
More accurate analysis of the ASCA spectra shows that:
 (1) The inner radius of the standard geometrically thin optically
thick part of the accretion disk exceeds $R_{in}>$10--25$R_g$.
 (2) The characteristic temperature in the inner part of the accretion flow
including the boundary layer exceeds $\sim$2 keV independently upon the
assumption about the density in the accretion flow. The electron scattering
gives the dominant contribution to the opacity in this region.
 
}                                                    
\bsk
\baselineskip = 12pt

\bsk
\baselineskip = 12pt


\text{\ni 1. INTRODUCTION
\ssk
\ni     
Observations of the X-ray bursters show that in the high luminosity state
($L_x> {\rm few} \times10^{37}$ erg/s) they have sufficiently soft spectra,
which can be often described as a superposition of a blackbody emission and an
emission of an optically thick accretion disk (e.g. White et al. 1988).
It is often assumed, that the blackbody component
originates in the optically thick boundary layer (BL) between the inner edge of
a Keplerian accretion disk and the surface of a slowly rotating neutron
star (Mitsuda et al. 1984, White et al. 1988). The importance of the BL
contribution was emphasized by many authors (e.g. Kluzniak\&Wagoner 1985,
Shakura\&Sunyaev 1986, White et al. 1988). In particular, it has been shown,
that even in Newtonian approximation the luminosity of the BL equals to the
luminosity of the extended accretion disk. 
In Schwarzschild geometry the luminosity of the BL can by a factor of 2
or more exceed that of accretion disk (Shakura\&Sunyaev 1986). As was shown

\newpage

\noindent
by Sibgatullin\&Sunyaev (1998) the account for the realistic time-space
geometry around a rotating neutron star reduces this
value for a prograde accretion disk, but for any reasonable parameters of the
rotating neutron star the BL luminosity exceeds that of the
accretion disk.
As was noted by White et al. (1988), the simplest approximation of the
spectra of X-ray bursters in the high state by the superposition of a
black body emission and an emission of an optically thick accretion disk
results in the blackbody luminosity less than predicted for the BL. 

In the low spectral state the spectra of X-ray bursters usually can be
adequately approximated by a single power law with low energy absorption and
are not consistent with significant contribution of the soft spectral
component to the total luminosity. Therefore assumption of an optically
thick BL emitting a nearly blackbody spectrum encounters even greater
difficulties. 

In this work we derive some simple constraints on the parameters
of BL and accretion disk basing on the fact that the spectra of X-ray
bursters in the low spectral state do not have strong soft components.

\begin{table}
\small
\caption{TABLE 1. The list of ASCA observations of several X-ray bursters
and best fit parameters of the power law approximation fo GIS2 and GIS3 data.\label{obslog}}
\begin{tabular}{lccccc}
\hline
X-ray&Date obs.&Exposure&$\alpha$&$N_H$&$L_x$, $\times10^{36}$erg/s$^a$\\
burster&       &sec     &        &$cm^{-2}$&0.5--10 keV\\
\hline
4U0614+091&18/04/93  &3548&$2.24\pm0.01$&$0.33\pm0.01$&$1.78\pm0.01$\\
4U1724-30&24/09/95   &8853&$1.77\pm0.02$&$1.60\pm0.03$&$3.60\pm0.04$\\
SLX 1735-269&15/03/95&18378&$2.13\pm0.01$&$1.45\pm0.02$&$1.75\pm0.02$\\
4U1850-56&06/10/95   &20444&$2.15\pm0.01$&$0.19\pm0.01$&$1.39\pm0.01$\\
\hline
\end{tabular}
\begin{list}{}{}
\item[$^a$] Luminosity not corrected for interstellar absorption, assuiming
2.5 kpc distance for 4U0614+091, 7.5 kpc for 4U1724-30, 8.5 kpc for SLX
1735-269 and 6.8 kpc for 4U1850-56. 
\end{list}
\end{table}



\ssk
\ni
2. CONSTRAINS ON THE PARAMETERS OF THE BOUNDARY LAYER AND ACCRETION DISK
\ssk
\ni
For our analisys we used data of ASCA (Tanaka, Inoue\&Holt 1994)
observations of four X-ray bursters in the low spectral state in 1993--1995. A
brief summary of the observations and the best fit parameters of the
power law with low energy absorption approximation are presented in Table 1.

A black body spectrum is a usual choice of a spectral model for a soft
component, in particular in the context of the BL emission (e.g. White et
al. 1988). It is well known however, that a black body emission spectrum
can be formed only if the free-free opacity dominates (see e.g.
Zeldovich\&Shakura 1969, Felten\&Rees 1972, Shakura 1972, Shakura\&Sunyaev 
1973). In the opposite case (scattering opacity dominates) scattering can
significantly distort the emergent spectrum. We shall assume for simplicity
that the region of the main energy release in the BL is isothermal and
consider two kinds of the density distribution -- exponential and homogeneous
(more realistic density profiles were considered in Illarionov\&Sunyaev, 1972).
The isothermal exponential atmosphere can be
formed near the surface of the NS if the local energy flux in the BL is
sufficiently smaller than the 
local Eddington flux. If the BL luminosity is $\sim10^{36}$ erg/s
and the emitting area of the BL is a small fraction of the neutron star
surface area ($<0.01$) then the local energy flux will be comparable with the
Eddington value and the radiation pressure will significantly modify the
density profile in the BL. In this case the spectrum of an isothermal
homogeneous atmosphere might be a better approximation to the BL spectrum.

For the spectral approximation of the ASCA data we used three different
two-component models: 1) power 
law + black body spectrum, 2) power law + spectrum of isothermal exponential
atmosphere and  3) power law + spectrum of isothermal homogeneous atmosphere.
 The scale height of isothermal exponential
atmosphere was calculated for any given temperature of the atmosphere
assuming a neutron star with mass 1.4 $M_{\odot}$ and radius 10 km. The
density and the height of
isothermal homogeneous atmosphere were fixed at $N=10^{22}$ cm$^{-3}$ and
$h=2000$ cm. The choice of density corresponds
to the minimal value required to produce luminosity $\sim 10^{36}$ erg/s
assuming that the emitting area equals to the neutron star surface area and
the temperature is $\sim$ 1 keV. The height of the atmosphere $h$ was chosen so
that the ``true opacity'' $\tau\sim\sqrt{\tau_{ff}(\tau_T+\tau_{ff})}>1$
for $kT=1$ keV at $E=1$ keV ($\tau_{ff}$ -- free-free, $\tau_T$ -- electron
scattering optical depth). In this case the spectrum is almost
independent on $h$ and depends only on the ratio 
$\tau_T/{\tau_{ff}}$ and as $\tau_T/{\tau_{ff}}\rightarrow 0$ ($\rho$
increases) the spectrum $\rightarrow$ blackbody spectrum. Fixing the
temperature of the soft component in the 0.2-2.0 keV interval and 
varying its normalization we obtained the upper limits on the soft component
luminosity as 
a function of temperature. In Fig.1 (left panel) we present the 2$\sigma$ upper
limits on the luminosity of the soft components for three spectral model
mentioned above as a function of $T$, in the units of total 0.5--10 keV
luminosity of the source. One can see that the contribution of the soft
component does not exceed $\sim$10--20\% of the 0.5--10 keV luminosity.
\footnote{Note, that for a typical low state spectrum the bolometric luminosity can
exceed the 0.5--10 keV luminosity by a factor of 2 or more.} Thus, the
contribution of the soft component is considerably lower than that expected
for the BL emission.

{\centering
\begin{figure}
\epsfxsize=11 cm

\epsffile[70 200 550 350]{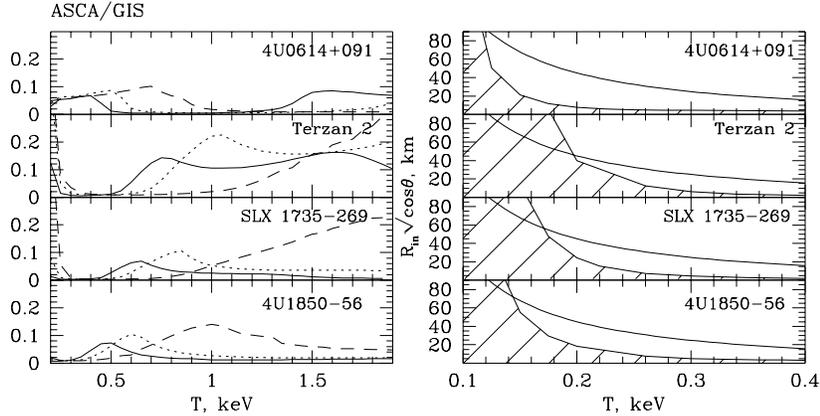}

\caption{FIGURE 1. Upper limits on the luminosity of the various soft
components in the units of total 0.5-10 keV flux (left panel). Solid line --
black body model, dotted line -- model of isothermal exponential,
dahsed line -- model of homogeneous atmosphere. In the right
panel the upper limits on the inner radius of the standard optically thick
part of the accretion disk as a function of its temperature are presented.
Solid line shows the theoretical dependence for parameter $\dot{m}=0.1$ .}
\end{figure}
}

Therefore {\em none of the considered simple models of the BL emission,
with the temperature in the range 0.2-2.0 keV, is consistent with the
ASCA data}. Moreover, as we considered low density (homogeneous
atmosphere with $N=10^{22}$ cm$^{-3}$) and high density (black body)
limits of the BL, we can tentatively conclude that {\em independently on the BL
density the BL temperature should exceed $\sim 2$ keV}. For such temperature
the emergent spectrum of BL should be significantly modified by the electron
scattering both in case of isothermal exponential and homogeneous atmosphere. 

The upper limits on the soft spectral component constrain also the
parameters of the optically thick part of the accretion disk.
Assuming that the emission of the optically thick part of the accretion disk
can be described by the simple multicolor black body disk model
(Shakura\&Sunyaev, 1973, Mitsuda et al., 1984) we can use the observational
data to obtain the upper limits on the inner radius of the  
accretion disc as a function of the disk temperature (Fig.1, right panel).
Furthermore, 
assuming that the 0.5--10 keV luminosity can be used to estimate the mass
accretion rate in the disk we can obtain the theoretical  
dependence of the temperature of the inner edge of the optically thick part
of the accretion disk on its radius.
The overlap of these two regions would define the allowed range of
$T_{in}$ and $R_{in}$. From the Fig.1 (right panel) one can see that {\em
the typical lower 
limit on the inner radius of the optically thick part of the accretion disk
is about 40--100 km. For a neutron star with $M=1.4M_{\odot}$ it corresponds to
$\sim$ 10--25$R_g$.} The regions closer to the compact object should have
spectra significantly modified by the electron scattering, so they will have
higher temperatures.

\bsk
\baselineskip = 12pt
{\abstract \ni ACKNOWLEDGMENTS
M.R. thanks the RBRF grant N980227566.Research has made use of data obtained
through the High Energy Astrophysics Science Archive Research Center Online
Service, provided by the NASA/Goddard Space Flight Center.

}

\bsk
\baselineskip = 12pt


{\references \ni REFERENCES
\ssk

\ref Czerny, Czerny\&Grindlay 1986
\ref Kluzniak\&Wagoner 1985
\ref Felten\&Rees 1972, A\&A, 17, 226
\ref Illarionov\&Sunyaev 1972, Astroph.\&Space Sci., 19, 61
\ref Mitsuda et al. 1984, PASJ, 36, 741
\ref Shakura 1972, Sv.A., v.16., No.3, p.532 
\ref Shakura\& Sunyaev 1986, PAZH, v.12, N.4, p. 286
\ref Sibgatullin\& Sunyaev 1998, in press, see also astro-ph/9811028
\ref Tanaka, Inoue\&Holt, 1994, PASJ, 46, L37
\ref White et al. 1988, ApJ, 324, 363
\ref Zel`dovich\&Shakura 1969, Astron. Zh., 46, 225 
}                      

\end{document}